\documentclass[prb,preprint,longbibliography]{revtex4-1} 

\usepackage{amsmath} 
\usepackage{longtable} 
\usepackage{graphicx} 
\usepackage{epstopdf} 

\begin{document}

\title{An epistemology and expectations survey about experimental physics: Development and initial results}

\author{Benjamin M. Zwickl}
\affiliation{School of Physics and Astronomy, Rochester Institute of Technology, Rochester, NY 14623}
\email{benjamin.m.zwickl@rit.edu} 
\author{Takako Hirokawa}
\affiliation{Department of Physics, University of Colorado Boulder, Boulder, CO 80309}
\author{Noah Finkelstein}
\affiliation{Department of Physics, University of Colorado Boulder, Boulder, CO 80309}
\author{H. J. Lewandowski}
\altaffiliation[Also at ]{JILA, University of Colorado Boulder, Boulder, CO 80309} 
\affiliation{Department of Physics, University of Colorado Boulder, Boulder, CO 80309}
\date{\today}

\begin{abstract}
In response to national calls to better align physics laboratory courses with the way physicists engage in research, we have developed an epistemology and expectations survey to assess how students perceive the nature of physics experiments in the contexts of laboratory courses and the professional research laboratory.  The Colorado Learning Attitudes about Science Survey for Experimental Physics (E-CLASS) evaluates students' epistemology at the beginning and end of a semester. Students respond to paired questions about how they personally perceive doing experiments in laboratory courses and how they perceive an experimental physicist might respond regarding their research.  Also, at the end of the semester, the E-CLASS assesses a third dimension of laboratory instruction, students' reflections on their course's expectations for earning a good grade.  By basing survey statements on widely embraced learning goals and common critiques of teaching labs, the E-CLASS serves as an assessment tool for lab courses across the undergraduate curriculum and as a tool for physics education research.  We present the development, evidence of validation, and initial formative assessment results from a sample that includes 45 classes at 20 institutions. We also discuss feedback from instructors and reflect on the challenges of large-scale online administration and distribution of results.  
\end{abstract}

\maketitle

\section{Introduction}

Laboratory courses offer significant opportunities for engagement in the practices and core ideas of science.  Laboratory course environments typically have apparatus, flexible classroom arrangements, low student/teacher ratios, and opportunities for collaborative work that promote students' engagement in a range of scientific practices (e.g., asking questions, designing and carrying out experiments, analyzing data, developing and refining models, and presenting results to peers).  Creating such opportunities requires significant investments in physical space, sophisticated equipment, and instructor support.  Despite the abundant opportunities and resources in many laboratory courses, concerns are frequently raised about how effective such courses are at fulfilling their potential.\cite{Trumper2003,Hofstein2004}  Problems often manifest themselves as a gap between the kinds of practices going on in the laboratory classroom and the practices going on in professional scientific research and engineering labs.  Sometimes gaps result from differing goals between lab courses and research experiences, while other times gaps result from good intentions, but poor implementation of the goals within the curriculum.  There are many calls to transform lab courses coming from the physics education community,\cite{Teachers1998} the life sciences,\cite{BIO20102003} and national science policies promoting the retention of STEM majors and the development of the STEM workforce.\cite{Singer2005,PCAST2012}  One theme that spans these calls is students should develop habits of mind, experimental strategies, enthusiasm, and confidence in research through effective laboratory courses.

A variety of responses have emerged for improving laboratory experiences within the physics curriculum.  Some laboratories have introduced new educational technologies (e.g., microcomputer-based labs\cite{Thornton1990} and VPython\cite{Buffler2008,Caballero2012}), others have added an emphasis on particular scientific practices (e.g., measurement and uncertainty\cite{Kung2005,Allie2003}, developing testable questions  and designing experiments \cite{Etkina2007a,Etkina2006b}, and scientific argumentation\cite{Moskovitz2011}), while others have pushed the lab course closer to cutting edge research by introducing modern physics concepts and apparatus (e.g., single photon quantum optics experiments\cite{Galvez2005,Pearson2010}), while others have demonstrated improved conceptual learning gains through research-based lab activities.\cite{Redish1997}  The diversity of responses reflects both the diversity of goals for the laboratory and the flexibility and adaptability of the laboratory environment  to meet many different goals.  Given this wide range of modifications to the laboratory curriculum,  there is a need for evaluation tools for lab courses that allow instructors to iteratively improve their course offerings, and for tools to  give physics education researchers insight into effects of different course modifications on student learning.  We have developed, validated, and collected initial results on a national-scale for a new epistemology and expectations (E\&E) survey\cite{Elby2011,Elby2001,Halloun1998,Redish1998,Lederman2002,Adams2006} called the Colorado Learning Attitudes about Science Survey for Experimental Physics (E-CLASS).\cite{Zwickl2012}${}^{,}$\footnote{URL: http://tinyurl.com/E-CLASS-Sp13-Post}  An E\&E survey is well-suited to assessing the present situation in laboratory instruction for four reasons. First, E\&E surveys are not directly tied to specific content, which increases their applicability in the already-existing wide range of laboratory courses.  Second, the habits of mind and ways of thinking probed in E\&E surveys represent a significant course goal for many instructors.  Third, in lecture courses, there is a demonstrated link between students' epistemology and their learning,\cite{Hammer1994,Lising2005} yet there is no epistemology assessment tool specifically designed for laboratory-centered instruction.  Fourth, E\&E surveys are of most value when evaluating educational environments that have significant differences from professional practice.  On the surface, lab classes have much in common with professional research (e.g., making predictions, carrying out experiments, analyzing data), yet the character of these activities may be significantly different in the two contexts.  This suggests lab courses may sometimes unintentionally confuse students' ideas about the nature of knowing and learning experimental physics.  
However, as lab courses are transformed to include more skills that prepare students for research, we expect gaps between students' and experts' epistemological beliefs about experiments will also narrow.  

The process for the development and validation of the E-CLASS as a course assessment tool for laboratory instruction broadly aligns with the procedures laid out in Adams and Wieman's article on the \textit{Development and Validation of Instruments to Measure Learning of Expert-Like Thinking},\cite{Adams2011} which aligns with the Standards for Psychological and Educational Assessment.\cite{AERA1999}  Our process begins with the identification of key areas of importance to instructors where students often differ from experts.  We then present our overall design criteria for the survey development.  Our development continues on to the validation and refinement of a ready-to-administer online assessment tool.  Initial results from the Fall 2012 and Spring 2013 semesters are presented as they appear in a typical post-semester report sent to instructors as a formative assessment tool.  We conclude by giving an overview of the level of participation across all classes, summarizing difficulties in achieving consistently high levels of participation, and looking at future research questions that can be answered using the E-CLASS.  

\section{Identifying differences between experts and novices in experimental physics}
\label{sec:Identifying_differences}
Like any tool for assessment of instruction, the E-CLASS must meet the triple criteria of (1) measuring something that experts and instructors care about (i.e., it should be aligned with widely accepted course goals), (2) targeting areas where students may not be meeting instructors' goals, and (3) accurately capturing some aspects of student thinking and learning. 

In order to measure something that most instructors care about, we aligned the survey with a set of consensus learning goals developed for our lab curriculum for physics majors,\cite{Zwickl2013} though there is considerable overlap with similar goals established by AAPT for the introductory labs.\cite{Teachers1998} Broadly, these goals were: modeling physical systems, modeling the measurement tools, statistical analysis, design of experiments and apparatus, troubleshooting, communication of scientific arguments, communicating in genres relevant to scientists, and technical lab skills using common lab equipment and software.  Beyond these learning goals that emerged through a departmental consensus-building process, we followed other E\&E surveys such as the Colorado Learning Attitudes about Science Survey (CLASS) by also considering students' affect and confidence when doing physics experiments and their identity as researchers.

In order to ensure the E-CLASS meets the second criteria of probing areas where students may not be meeting instructors' goals, we aligned the survey with several common challenges that instructors have found in our lab courses at the University of Colorado Boulder and are common elsewhere.  We knew many students found the labs very time-consuming and many students disliked our introductory lab course.  Does this impact their general enthusiasm for doing experiments in a research setting?  Students repeat historic experiments with known results rather than asking their own questions  and designing experiments to investigate them.  Does this impact how they think about the roles of asking questions, design, and confirmation in conducting research?  Students are often confronted with a range of new measurement tools and apparatus.  Do our students treat the apparatus as something to be understood and explored or as a ``black box''?  Uncertainty analysis and error propagation has played a significant role in our curriculum as well.  Do our students see uncertainty as a tool for better understanding their data and refining their experiment, or is it just an algorithmic calculation that comes at the end of the lab report?   As the final step of most of our lab activities, students complete a lengthy written report that often takes more time to complete than they spend working with the equipment and taking data.  Do students see writing lab reports as an exercise in scientific communication or merely in meeting the instructor's grading expectations?  For fear of cheating in our large introductory course, students have often been required to work individually in the lab.  When students work by themselves, does it  affect the role they see for collaboration within scientific research or lessen the value they place on presenting ideas to peers?  These kinds of concerns helped us target the E-CLASS statements on areas where we may see larger signal and provide relevant information for formative assessment.  

The final criteria, that the E-CLASS should accurately capture some aspects of students' thinking and learning, is explored in the following sections as we articulate more clearly what is probed (Sec.\ \ref{sec:Design_considerations}), and then present evidence of validity (Sec.\ \ref{sec:validation}).

\section{Survey design considerations}
\label{sec:Design_considerations}
\subsection{Measuring epistemology and expectations in the lab}

The E-CLASS was designed to survey students' \textit{epistemological} beliefs and their \textit{expectations}.  \textit{Epistemology} refers to theories of the nature of knowledge, knowing, and learning in the discipline.\cite{Kuhn2000,Hofer1997,Elby2009}   Epistemology, in the context of the lab, means defining what is viewed as a good or valid experiment and what are the appropriate ways to understand the design and operation of an experiment and the communication of results.  The E-CLASS also includes students' views about learning experimental physics as part of their overall epistemology.\cite{Elby2009}  \textit{Expectations}, on the other hand, deal with students' perceptions of what their instructor expects they should be doing in the class---the kinds of knowledge and learning that are expected and rewarded in the laboratory course.  While expectations are often evaluated at the beginning of the course, we included reflective questions about the course's expectations as part of the post-survey. We believe such reflections (which form a triplet with the personal and professional epistemology statements) give more direct feedback to the instructor and are something an instructor can influence through explicit framing, grading priorities, and classroom culture.  In order to assess the impact of the course, the E-CLASS provides pre and post measures of students' personal and professional epistemology, while also providing a post-only reflective look at expectations.   Personal and professional epistemology questions are always presented as a pair, and when appropriate a third question is added about expectations.  In the post survey, 23 of 30 statements are associated with the triplet of epistemology and expectations questions, while the remaining 7 are only personal and professional epistemology pairs (see Appendix for the full list of statements).   The inclusion of linked epistemology and expectations questions allows E-CLASS to directly evaluate relationships between epistemology and expectations in the course.

As a course assessment tool, we wanted to cover many important aspects of experimental physics.  Probing a wide range of epistemological statements allows the survey to have relevance in courses that have a wide range of goals.  We also take a \textit{resources} perspective\cite{Hammer2005,Louca2010,Yerdelen-Damar2012} on the nature of these epistemological beliefs.  This means that we don't expect students to hold particularly coherent epistemological stances as though they had some well-developed world-view of doing physics experiments.  Instead, we expect students to draw on a range of resources and experiences when responding to each statement, and responses might sometimes be in apparent contradiction with each other due to contextual differences (e.g., Sec.\ \ref{sec:Personal_Prof_Splits} shows an apparent contradiction in students' epistemology about the role of experiments for generating new knowledge).\cite{Yerdelen-Damar2012}    Because of this resources perspective, we do not use the survey as a tool to evaluate individual students, but as a coarse measurement of the epistemological state of the class.\cite{Elby2011}  

\subsection{Format and structure of the survey}

For ease of administration, we followed the Colorado Learning Attitudes about Science Survey in the use of Likert-scale responses to statements.\cite{Adams2006}  However, unlike the CLASS, we did not develop categories for clustering questions in the data analysis.  The E-CLASS questions still form groups that align with the course goals described in Sec.\ \ref{sec:Identifying_differences}, but those groups were not used to create statistically robust categories via factor analysis. The reasons for the omission of categories are two-fold.  The first deals with the nature of actual lab courses.  It is possible for a course to prioritize one idea, while ignoring another related idea.  In other words, the category may make sense to an expert, but the correlation may not be reflected in students' responses.  For instance, it seems reasonable that ``communicating results using scientific arguments'' and ``communicating scientific results to peers'' are ideas that could be grouped in the same ``communication''  category.  Yet often courses are structured so that students' results are only communicated to the instructor, while communication to peers is ignored as a course goal.  So although audience and argumentation are each aspects of communication, they can be emphasized independently in a course. The second reason for omitting categories is that our standard presentation of results was primarily designed to give actionable feedback that instructors could use to improve their courses.  By compactly presenting the results from each statement and sorting results from highest to lowest fractional agreement with experts (see Fig.\ \ref{fig:Personal_pre_post_shifts}), instructors can quickly identify items of most concern and start to consider aspects of their course that may influence this area of experimental epistemology. Although categories were not used in this initial version of E-CLASS, they may increase the survey's utility for broadly contrasting the epistemological impact of different curricular approaches.  The introduction of categories will be reconsidered in future versions of the survey. 

\section{Iterative validation and refinement of the survey}
\label{sec:validation}
\subsection{Lessons from early student interviews}

The initial development of the E-CLASS survey was based closely on the well-studied CLASS survey\cite{Adams2006} that has found significant use in undergraduate physics courses both in the United States and internationally.\cite{Alhadlaq2009,Zhang2013a}  We had reason to believe a straightforward adaptation might be possible.  A similar process of adapting CLASS from physics to chemistry\cite{Adams2008} was accomplished through a straightforward modification of many statements by changing the word ``physics'' to ``chemistry,'' by focusing attention on chemical formulas in addition to mathematical formulas, and by adding 11 new statements involving chemistry-specific ideas.  Validation interviews and faculty surveys for the CLASS-Chem showed the survey had a similar level of expert and student validity as the original CLASS-Phys.  We developed our own minimal adaptation by replacing many uses of the word ``physics'' or ``physics problem'' with ``experiment,'' and developing several new questions as well.  But in a series of 11 student validation interviews in Fall 2011, a substantial number of issues arose.    One of the most significant issues was that the phrase ``physics experiment'' is used to refer to activities in a lab class \textit{and} to the kinds of experiments that professional researchers engage in.  Depending on the exact statement, students switched between a context of classroom laboratories, everyday life, and professional physics experiments, and their answers could depend very strongly on which context they chose.  In addition, students often commented that they were unsure whether they should answer ``What do I think?'' or ``What should I think?'' when asked to rate their level of agreement about a statement like ``When doing an experiment, I just follow the instructions without thinking about their purpose.''  The final difficulty of this early version of the survey was that it did not probe many aspects of experimental physics that we viewed as important (i.e., it was too disconnected from our learning goals).

Because of these early interviews and a desire to more strongly link to the consensus learning goals, the later iterations of E-CLASS began to differ more significantly from CLASS-Phys.  The primary changes were (1) that we focused the context of students' responses to either be about ``experiments for class'' or about ``research,''  (2) we also eliminated confusion about ``should I...'' vs ``do I...'' by asking students paired questions that distinguished between what they thought and what an expert might think, similar to the paired format using the FCI\cite{McCaskey2004a,McCaskey2004} and CLASS-Phys,\cite{Gray2008a} and (3) the statements in the survey more effectively spanned our assessment goals described in Sec.\ \ref{sec:Identifying_differences}.  

\subsection{Creation of a final version} 
\begin{figure}
\includegraphics[width=0.45\textwidth, clip, trim=0mm 0mm 0mm 0mm]{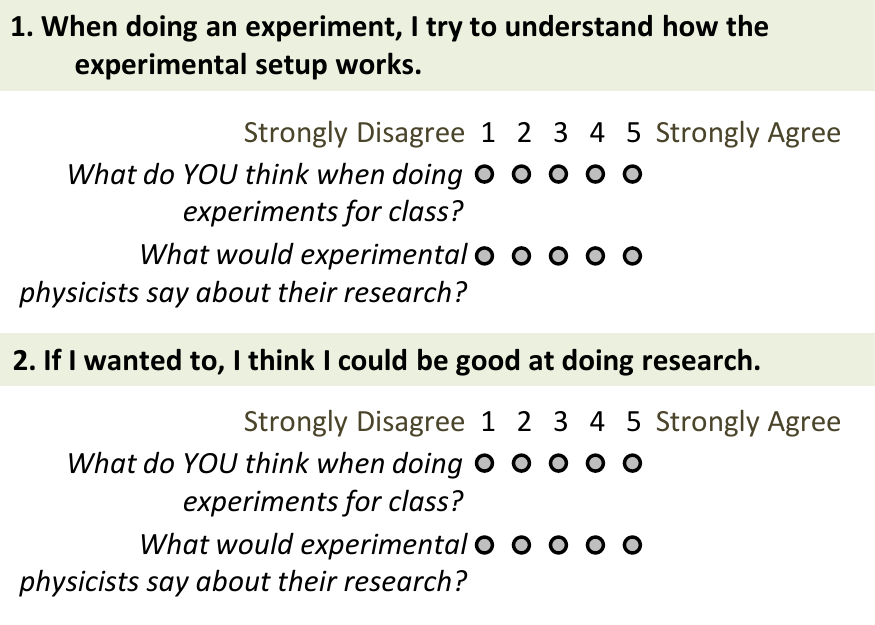}
\caption{Two epistemological beliefs statements as they are presented to students in the pre- and post- E-CLASS online survey.}
\label{fig:Pre_Post_Questions_Example}
\end{figure}

\begin{figure}
\includegraphics[width=0.45\textwidth, clip, trim=0mm 0mm 0mm 0mm]{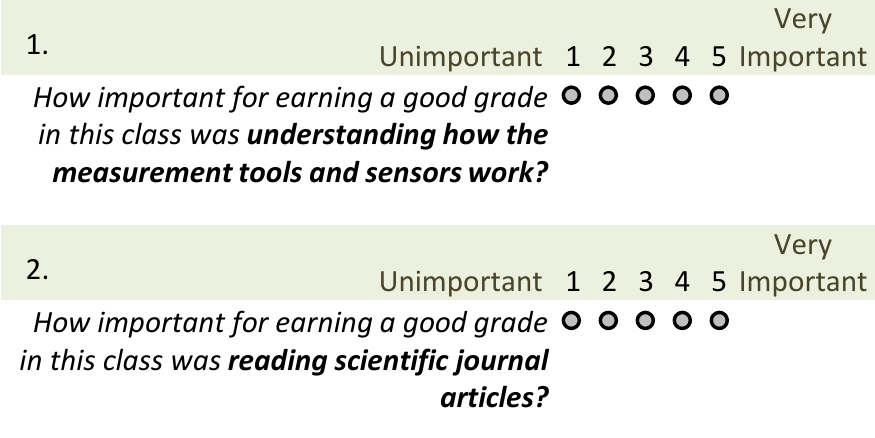}
\caption{Two expectations questions as they are presented to students in the post-semester E-CLASS online survey.}
\label{fig:Post_Questions_Example}
\end{figure}

Figures \ref{fig:Pre_Post_Questions_Example} and \ref{fig:Post_Questions_Example} show a few statements from the E-CLASS in the format they were presented to students in the online surveys during Fall 2012 and Spring 2013.  All thirty pairs of personal and professional epistemology statements are presented in a format similar to Fig.\ \ref{fig:Pre_Post_Questions_Example}.  The subset of 23 (out of 30) statements that have a corresponding expectations statement are presented in a separate section at the end of the post survey in a format similar to Fig.\ \ref{fig:Post_Questions_Example}.  In order to come to this final format and question wording, thirty one additional interviews were conducted during Spring and Fall 2012.  These interviews were focused on three aspects of the survey design.  One aspect was refining the question wording to clarify the context of students' epistemological reflections.  Through these interviews, the paired survey questions evolved from ``What do YOU think?'' and ``What would a physicist say?'', which were used in the paired CLASS-Phys\cite{Gray2008a}, toward the current wording ``What do YOU think when doing experiments for class?'' and ``What would experimental physicists say about their research?''  The second emphasis of the interviews was on the wording of individual statements to make sure they could be readily interpreted by all levels of undergraduate students.  The third focus was on how students interpreted the phrase ``experimental physicists'' and whether that could be replaced with more general language of ``scientists.''  We discuss each of these aspect in turn.

\subsection{Evidence of validity through student interviews}

In order to ensure reliable interpretation of the context for students' responses to the epistemology statements, we found it necessary to clarify ``What do YOU think?'', which was the prompt used in the paired CLASS-Phys.  Most frequently, students answered ``What do YOU think?'' by reflecting on their prior experience in lab classes, but students with prior research experience, especially upper-division students, often would switch to a context of their own personal research experience if it seemed to fit.  The final wording ``What do YOU think when doing experiments for class?'' ensures students maintain a consistent context for reflecting on ``doing physics experiments.''  This wording also aligns with the default interpretation of students who have never taken a physics lab prior to taking the E-CLASS.  Such students typically referred to their experiences in a high school or introductory college-level science lab or to in-class demonstrations that involved observations of real phenomena.

The question ``What would experimental physicists say about their research?'' also went through successive iterations.  Because experiments exist in very different forms in both research and teaching labs, and because professional physicists participate in both of those environments (as either teachers in teaching labs or researchers in research labs), we restricted the context of the question to research.  The use of ``experimental physicists'' rather than ``physicists'' arose in the interviews to clarify the question for upper-division students who are becoming more aware of the professional culture within physics.  In the interviews, it was suggested theoretical physicists may hold differing views, particularly regarding statements about enjoyment while doing experiments or building things and working with their hands.  

In summary, the use of the paired statements ``What do YOU think when doing experiments for class?'' and ``What would experimental physicists say about their research?'' clarifies students' interpretation of the questions and also clarifies the meaning of the E-CLASS as an assessment tool.  In this final form, the E-CLASS measures students epistemological beliefs about physics experiments in the two contexts where such experiments occur: in the laboratory classroom and in the research lab.  The E-CLASS becomes a tool to assess students' perception of the gap between their own classroom experiences and what they perceive to be the nature of authentic science.  While the E-CLASS uses pairs of statements in two different contexts for the reasons stated above, it does differ from the paired CLASS-Phys\cite{Gray2008a}, which used the same general context (neither classroom nor research) for evaluating students' views of personal and expert epistemology. 

The second focus of these validation interviews was to refine individual question wording.  For instance, in a trial administration of the E-CLASS in Spring 2012 to introductory students at CU, we found that the majority of students agreed with the statement ``I can read a journal article for understanding.''  Given the difficulty of reading the professional literature for graduate students, we were surprised that so many introductory students would agree with this statement.  Through targeted validation interviews, we found that many students set a low bar for ``reading for understanding'' that was equated with ``getting the gist of the article.''  Also when discussing ``journal articles,'' some introductory students mentioned popular science articles (e.g., Scientific American) rather than professional research literature that was intended in our question.  The final question wording was substantially modified to address these findings and now reads: ``Scientific journal articles are helpful for answering my own questions and designing experiments.''  For other statements, particular attention was paid to technical terms, and where appropriate, question wordings were simplified.  For example, ``Doing error analysis (such as calculating the propagated error) usually helps me understand my results better.'' was simplified to ``Calculating uncertainties usually helps me understand my results better.''  Not all technical language was avoided, but it was simplified whenever possible.  The remaining technical terminology was retained so that the survey would continue to address key aspects of experimentation for upper-division physics majors. 

The third aspect of the interviews dealt with the concern of some instructors that most introductory physics courses primarily serve non-physics majors, and the  use of the phrase ``experimental physicists'' makes an unhelpful distinction between experimental physicists and other scientists and engineers.  In particular, some faculty were worried the language may alienate students who are not physics majors by suggesting the material is only relevant to this small group of people called ``experimental physicists.''  A final series of interviews was conducted to better understand what comes to mind when students think about ``physicists'' and ``experimental physicists.''  The outcome was that most students were more aware of physicists famous for their theoretical ideas (e.g., Newton and Einstein), and had trouble naming any experimental physicists.  In addition, many introductory-level students were unfamiliar with the distinctions of theorist and experimentalist, but they still interpreted ``experimental physicists'' straightforwardly as ``physicists who do experiments.''  So the clarification does not obscure students' interpretation, but may help depending on whether a student is aware of the broader community of professional physicists.  We also investigated replacing the term ``experimental physicists'' with ``scientists.'' In interviews, students found ``scientists'' too general to answer the questions because they realized that scientists' typical activities could differ substantially between disciplines (e.g., an experimental physicists versus a field ecologist).   Lastly, even though the context was specific to experimental physics, most students still felt that the statements emphasized broadly relevant experimental skills that could be applied to their own discipline.

In order to gather evidence of validity across the broad population of students taking physics laboratory courses, altogether 42 interviews were conducted.  There were 24 students interviewed who had never taken any college physics lab classes, 8 were currently enrolled in an introductory physics lab, and the remaining 10 were physics majors who had already taken upper-division physics lab classes.  The high representation of non-physics majors in the validation interviews was needed because enrollments in introductory courses are typically dominated by students from outside of physics. The pre-introductory and introductory-level students included a mix of physical science majors, life science majors, and engineering majors.  The population of 42 interviewees included 22 male and 20 female. 

\subsection{Content validity}

Another key aspect of developing an assessment tool around epistemology is ensuring faculty have consistent responses to the various survey items.  We establish the \textit{content validity} of the E-CLASS when experts find the questions relevant (as described in Sec.\ \ref{sec:Identifying_differences}) and have a consistent response to the statements.  To date, we have collected 23 expert responses (3 full time instructors and 20 with a blend of teaching and on-going research in experimental physics) from both primarily undergraduate serving institutions ($N=7$) and PhD granting institutions ($N=16$).   Faculty were asked to respond to the thirty statements on the epistemological portion of the survey considering their own perspective as a faculty member and/or researcher.  In these responses, 24 of the 30 statements had an expert consensus of 90\% or higher, and all 30 statements had consensus of 70\% or higher.  The statements and distribution of responses with lower than 90\% consensus are summarized in Table\ \ref{tab:Low_expert_consensus}.

Despite the fact that a few of the questions had lower levels of expert consensus, we justify the inclusion of these statements for the following reasons.  The first three statements in Table\ \ref{tab:Low_expert_consensus} all relate to key learning goals of many labs: developing scientific arguments based on data, evaluating uncertainty in data, and understanding the theoretical ideas underlying the lab.  Although there was some small disagreement of the importance of these, they still remain important in many lab curricula and in the research programs of many faculty.  The fourth statement, about asking help from an expert, has an awkward context in the faculty survey, but we left the statement in for completeness as it does have a clear meaning in a classroom context for students.  Perhaps the most surprising and interesting results from the expert validation are two statements with the lowest consensus.  Over 25\% of respondents did not agree that working in a group is an important part of doing physics experiments, which might indicate that faculty have a variety of ways in which they go about their research depending on their particular research expertise and nature of their projects.  We retain this question because group work is typically an attribute of authentic research and also because  there are many pedagogical benefits to working in groups.  Finally, responses to the statement with the lowest consensus showed that about 30\% of instructors did not agree that nearly all students are capable of doing a physics experiment if they work at it.  This finding seems to indicate that faculty, when reflecting on their role as researchers, think physics experiments are difficult.  Most research faculty have many stories to tell of highly qualified students struggling in the lab, so perhaps their own experience suggests not \textit{all} students are capable of doing PhD-level experiments.  We retain this statement because we want to know whether students view physics experiments as accessible to a broad population.  A key motivation for improving laboratory instruction is improving retention in STEM, so it is critical that students see technically challenging aspects of STEM, such as doing physics experiments, as something accessible to many people.  

\begin{table*}
    \caption{A list of E-CLASS statements with the faculty agreement less that 90\%.  \textbf{Agree} is the number of respondents who answered either ``Agree'' or ``Strongly Agree''.  \textbf{Disagree} is the number of respondents who answered either ``Disagree'' or ``Strongly Disagree''.  \textbf{Consensus} refers to the fraction of respondents in the consensus response.}
    \begin{tabular}{|p{0.55\textwidth}| r|  r |r | r|}
        \hline
	\textbf{Statement} & \textbf{Agree} & \textbf{Neutral} & \textbf{Disagree} & \textbf{Consensus}\\ \hline\hline
	If I am communicating results from an experiment, my main goal is to make conclusions based on my data.  & 20 & 2 & 1 & 0.87 \\ \hline
	Calculating uncertainties usually helps me understand my results better. & 19 & 2 & 2 &0.83 \\ \hline
	I am usually able to complete an experiment without understanding the equations and physics ideas that describe the system I am investigating.   & 0 & 4 & 19 & 0.83 \\ \hline
	When I encounter difficulties in the lab, my first step is to ask an expert, like the instructor.   & 0 & 5 & 18 & 0.78 \\ \hline
	Working in a group is an important part of doing physics experiments.   & 17 & 4 & 2 & 0.74 \\ \hline
	Nearly all students are capable of doing a physics experiment if they work at it.   & 16 & 4 & 3 & 0.70 \\ \hline
          \end{tabular}
    \label{tab:Low_expert_consensus}
\end{table*}

\subsection{Convergent validity}

Evidence of convergent validity of an assessment tool shows that the assessment results are correlated with other established measures, such as course performance or GPA.  On similar assessment tools, such as the CLASS, it is found that students with more expert-like perspectives on physics and learning physics tend to do better in their physics courses.\cite{Perkins2005a}  To date, we have not had access to course grade data to correlate with E-CLASS scores, though we plan to do this analysis in upcoming semesters.  On the other hand, our current data set does contain a student population that includes many introductory-level non-physics majors and upper-division physics majors.  We expect that students who are majoring in physics and are taking upper-division labs would tend to have more expert-like views.   When comparing students in algebra-based physics labs to students in upper-division labs and averaging across all 30 statements, we find that upper-division students have a larger fraction of expert-like responses in both the classroom context (mean expert-like fraction = 0.66 vs 0.61, $p$-value = $6\times10^{-6}$, Cohen's $d$ effect size = 0.38) and in the context of professional research (mean expert-like fraction = 0.82 vs 0.78, $p$-value = $2\times10^{-4}$, Cohen's $d$ effect size = 0.28).  While the effect sizes reported are not large, upper-division students tended to be more articulate when explaining their responses during the validation interviews, so there is likely additional growth in epistemological sophistication that is not fully captured by aggregated E-CLASS scores.  This would suggest some higher-level epistemology statements should be added to the survey. 

At this point it is worth clarifying the valid use of E-CLASS across the undergraduate curriculum.  Student interviews reveal that the survey has a consistent interpretation across levels, meaning the pre/post results from an individual class can be meaningful for introductory through upper-division classes.  However, because interviews revealed greater differences in epistemological sophistication than was indicated by Likert-scale responses, any comparisons between different levels of courses should be limited until higher-level questions are added to future versions of E-CLASS and additional validity studies are performed.

\section{E-CLASS results as a course assessment tool}

\begin{figure}
\includegraphics[width=0.40\textwidth, clip, trim=0mm 0mm 0mm 0mm]{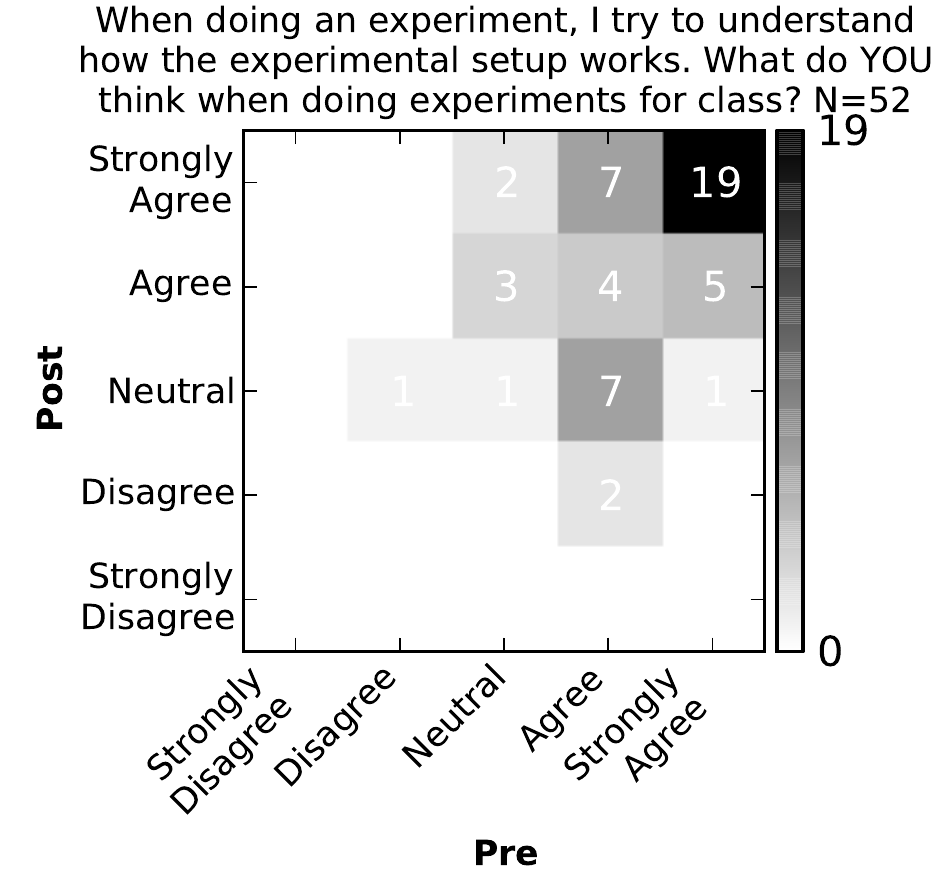}
\caption{The pre and post response data for a single statement summarized as a 2D histogram.  The number inside each box corresponds to the number of students with each (pre,post) response.}
\label{fig:Histogram_2D}
\end{figure}

\begin{figure*}
\includegraphics[width=0.70\textwidth, clip, trim=0mm 0mm 0mm 0mm]{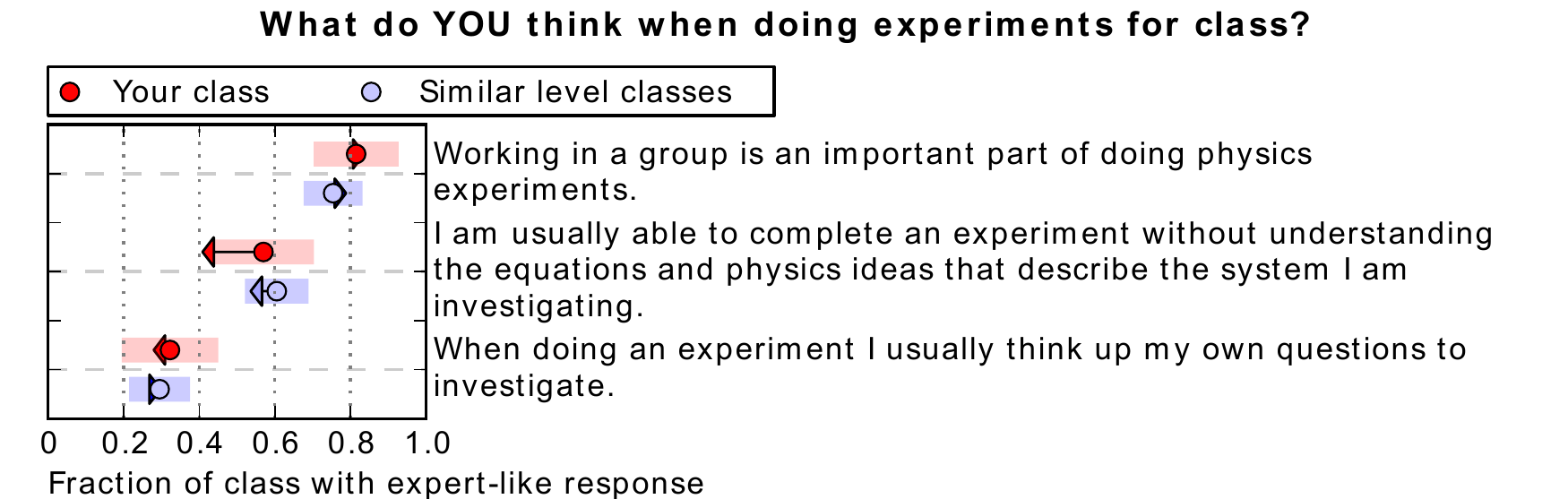}
\caption{Pre/Post changes in students' personal views about \textit{``What do YOU think when doing experiments for class?''} for \textbf{your class} (Red) and all students in \textbf{similar level classes} (i.e., introductory calculus-based physics classes) (Blue).  The circles show the pre-semester survey values. The arrows indicate the pre/post changes. The shaded bars are 95\% confidence intervals.  The data shown are for a subset of 3 out of 30 statements.}
\label{fig:Personal_pre_post_shifts}
\end{figure*}
The E-CLASS was designed with two purposes in mind.  The first purpose was as an assessment tool for laboratory courses.  The second was as a physics education research tool.  The results that follow demonstrate how the E-CLASS has been used as a course assessment tool for 45 classes during the Fall 2012 and Spring 2013 semesters.  We postpone the discussion of E-CLASS as a PER tool for comparative evaluation of different laboratory curricula for a later publication.  

One significant feature of the E-CLASS is that at the end of the semester, instructors are provided with a detailed results report in PDF format with explanations, analysis, and figures.  Figs.\ \ref{fig:Histogram_2D}, \ref{fig:Personal_pre_post_shifts}, \ref{fig:Overall_ECLASS_score}, \ref{fig:Grade_plot}, \ref{fig:Personal_professional_split}, and \ref{fig:Change_in_interest} and Table\ \ref{tab:Participation_Summary} are all full or abbreviated versions of those appearing in the instructor report for the lab component of an introductory calculus-based course at a large university (not CU-Boulder).  

\subsection{Personal epistemology}

\begin{figure}
\includegraphics[width=0.40\textwidth, clip, trim=0mm 0mm 0mm 0mm]{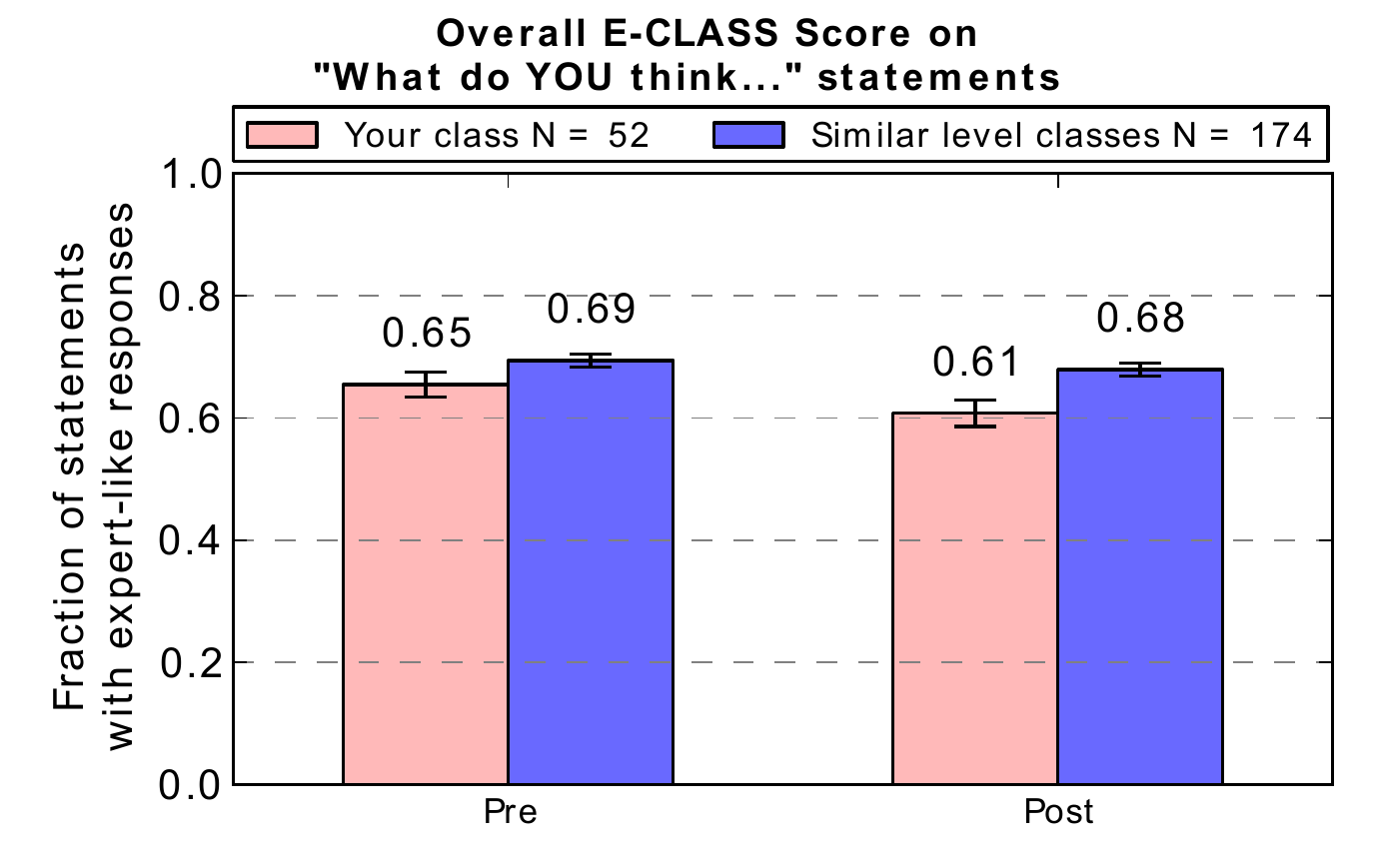}
\caption{Comparison between overall pre and post scores for students' personal views about \textit{``What do YOU think when doing experiments for class?''} \textbf{Your class} (Red) is compared with all students in \textbf{similar level classes} (i.e., introductory calculus-based physics classes) (Blue). The error bars show 95\% confidence intervals. The overall mean shown here averages over all students and all statements on the survey.}
\label{fig:Overall_ECLASS_score}
\end{figure}

The report begins by using one of the 30 questions as an example for how the pre/post shifts are calculated.  Fig.\ \ref{fig:Histogram_2D} shows the combined (pre,post) data for a single statement.  This information is then condensed to a pair of numbers---the fraction of student responses in agreement with experts on the pre-semester survey and on the post-semester survey.  That pair of numbers is used to generate plots of pre/post shifts as shown in Fig.\ \ref{fig:Personal_pre_post_shifts}.   Although Fig.\ \ref{fig:Personal_pre_post_shifts} shows pre/post shifts for only three statements, instructors receive a full version with all 30 statements.  Finally, the pre and post results from all questions can be further condensed into a single overall pre and post score for the class, as shown in Fig.\ \ref{fig:Overall_ECLASS_score}.  
Whenever possible, we also provide a comparison with a group of students in similar level classes.  The comparison group provides instructors with a baseline for evaluating whether or not their results are typical.  Currently, we are using three comparison groups: non-calculus-based introductory physics, calculus-based introductory physics, or upper-division (anything after the introductory level).  

\begin{table}
    \caption{Summary of class participation for an introductory calculus-based physics lab class at a large university.}
    \begin{tabular}{|p{0.35\textwidth}| r|}
        \hline
Number of valid pre-responses &	69 \\ \hline
Number of valid post-responses &	65 \\ \hline
Number of matched responses & 52 \\ \hline
Reported number of students in class &	117 \\ \hline
Fraction of class participating in pre and post	& 0.44 \\ \hline
          \end{tabular}
    \label{tab:Participation_Summary}
\end{table}

\subsection{Expectations}

\begin{figure*}
\includegraphics[width=0.70\textwidth, clip, trim=0mm 0mm 0mm 0mm]{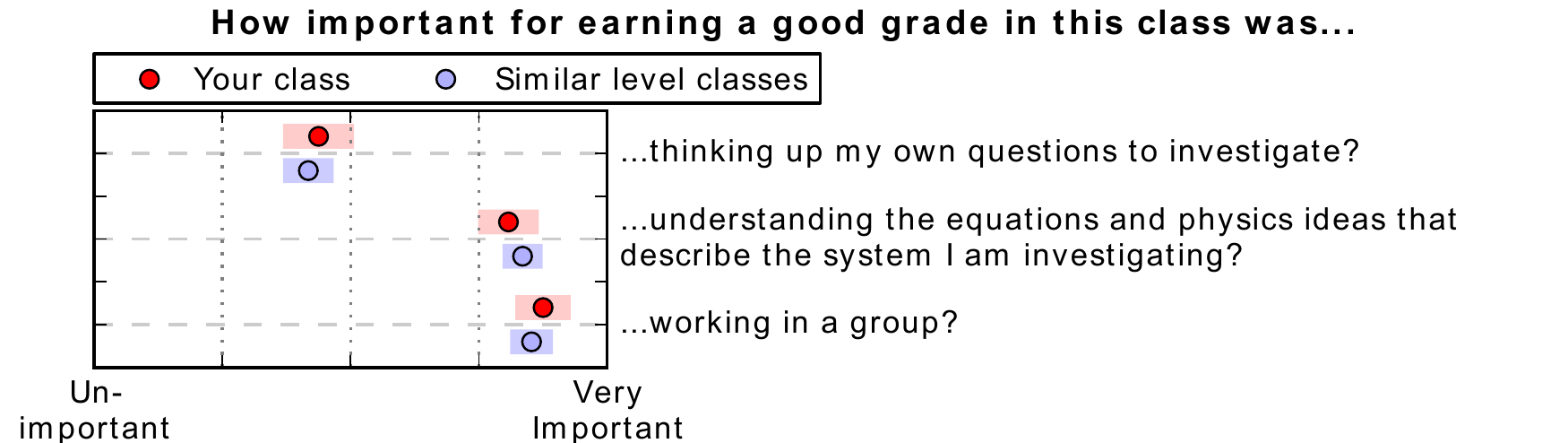}
\caption{Students' views of the importance of different activities for earning a good grade in \textbf{your class} (Red) and in \textbf{similar level classes} (i.e., introductory calculus-based physics classes) (Blue).  The data shown are for a subset of 3 out of 23 statements.}
\label{fig:Grade_plot}
\end{figure*}

The results discussed so far in Figs.\ \ref{fig:Histogram_2D}, \ref{fig:Personal_pre_post_shifts} and \ref{fig:Overall_ECLASS_score} deal only with students' responses to ``What do YOU think when doing experiments for class?'', which is just one part of the triplet of statements surrounding a single idea.  The second aspect of the E-CLASS survey is students' views of what was expected of students for earning a good grade.  The results of ``How important for earning a good grade in this class was...'' are shown in Fig.\ \ref{fig:Grade_plot}.  Such a plot allows instructors to see whether students' perceptions of the grading priorities for the class actually align with their own personal goals as instructors.

\subsection{Personal and professional splits}
\label{sec:Personal_Prof_Splits}

\begin{figure*}
\includegraphics[width=0.70\textwidth, clip, trim=0mm 0mm 0mm 0mm]{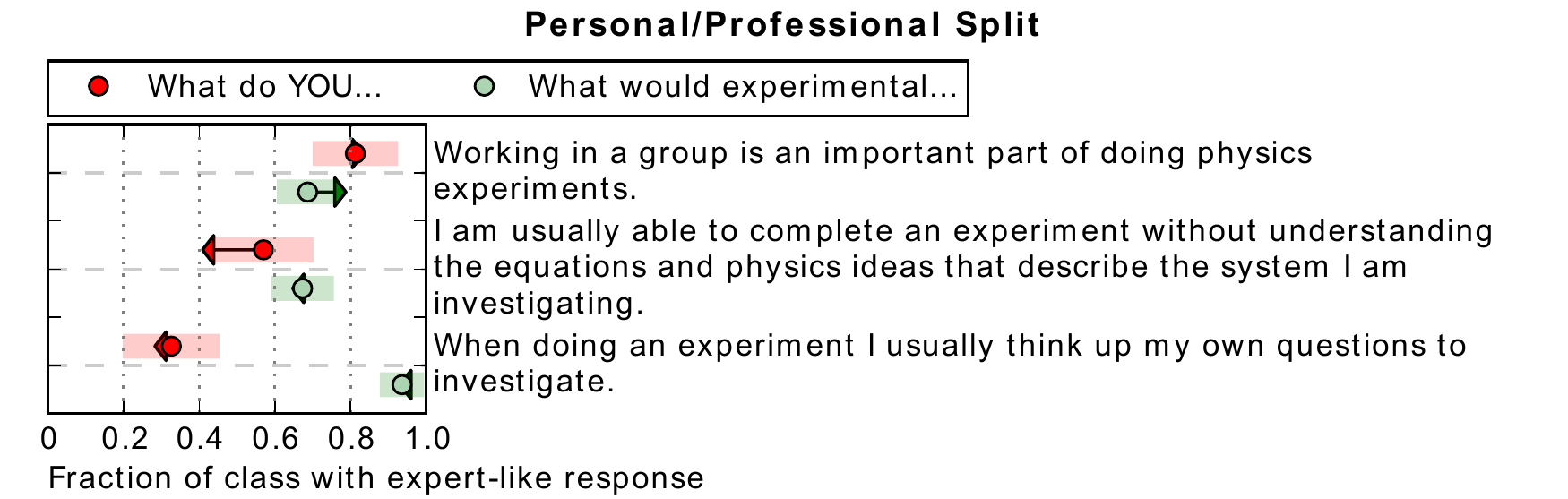}
\caption{Comparison of changes in students' personal views versus their views about professional physicists. Red shows the change in students' response to \textit{``What do YOU think when doing experiments for class?''}  Green shows the change in students' responses to \textit{``What would experimental physicists say about their research?''}  The circles show the pre-semester survey values. The arrows indicate the pre/post shift. The shaded bars are 95\% confidence intervals.  The data shown are for a subset of 3 out of 30 statements.}
\label{fig:Personal_professional_split}
\end{figure*}

The third area probed by the E-CLASS is students' epistemology regarding physics experiments done for research.  Data for this aspect of students' epistemology are shown in green in Fig.\ \ref{fig:Personal_professional_split}.  Although the data shown are for a subset of 3 of the 30 questions,  we typically find that students have much more expert-like views regarding what experimental physicists would say about their research than they do for their personal views about doing experiments for class.  We also find that students' views of researchers typically change less during the semester than their personal views.  One immediate use of these data is to identify the statements with the largest epistemological splits between students' views of classroom experiments and research experiments. For this particular class, the largest split occurred for ``When doing an experiment, I usually think up my own questions to investigate.''   (See Fig.\ \ref{fig:Personal_professional_split})  About 28\% personally agreed when thinking about experiments done for class, while 90\% thought experimental physicists doing research would agree.  Other questions with large splits (a difference of 40\% or larger) were: ``I don't enjoy doing physics experiments.'' ``Calculating uncertainties usually helps me understand my results better.''   ``Scientific journal articles are helpful for answering my own questions and designing experiments.'' and  ``If I don't have clear directions for analyzing data, I am not sure how to choose an appropriate analysis method.''

We can also use the data presented in Fig.\ \ref{fig:Personal_professional_split} to identify statements where students express the least agreement with experts' views about professional research.  For this class of calculus-based physics students, when asked ``What would experimental physicists say...,'' about 5\% of students disagreed with the statement, ``If I am communicating results from an experiment, my main goal is to create a report with the correct sections and formatting.'' When faculty were given same statement,  96\% disagreed, nearly the opposite result from students.  This results persists across many classes.  Among the 612 responses in the Spring 2013, 13\% of all responses disagreed.  Upper-division classes had disagree fractions as high as 40\% demonstrating upper-division students tended to have more expert-like views.  However, the divide between students and experts is so striking that we plan to conduct follow-up interviews to see what students are attending to and how it might differ from experts.  One hypothesis based on our own experience teaching lab courses is that an overemphasis on well-formatted lab reports may be misrepresenting the priorities of scientific communication.\cite{Allie1997,Moskovitz2011}  The statement with the second least expert-like result is ``The primary purpose of doing a physics experiment is to confirm previously known results.''  Only about 40\% of students disagreed when asked ``What would experimental physicists say...,'' while 100\% of experts disagreed.  This response is in apparent contradiction with the result that 94\% of students in the same class agreed with the statement ``Physics experiments contribute to the growth of scientific knowledge.''  This contradiction between two similar items extends beyond this class and is robust across a wide population of students and courses.  We plan to conduct a follow-up study to locate the source of the source of the contradiction, but from a resources perspective, it could be that subtle contextual features of the statements are triggering different epistemological resources.\cite{Yerdelen-Damar2012}

\subsection{Course participation}
In addition to summarizing the class' responses to each individual statement and question, we also provide instructors with a summary of their students' participation in the E-CLASS survey (Table\ \ref{tab:Participation_Summary}).  The classroom participation data  shown in Table \ref{tab:Participation_Summary} apply to the figures presented in Figs.\ \ref{fig:Histogram_2D}--\ref{fig:Change_in_interest}.

\subsection{Demographics and other information}
\begin{figure}
\includegraphics[width=0.40\textwidth, clip, trim=0mm 0mm 0mm 0mm]{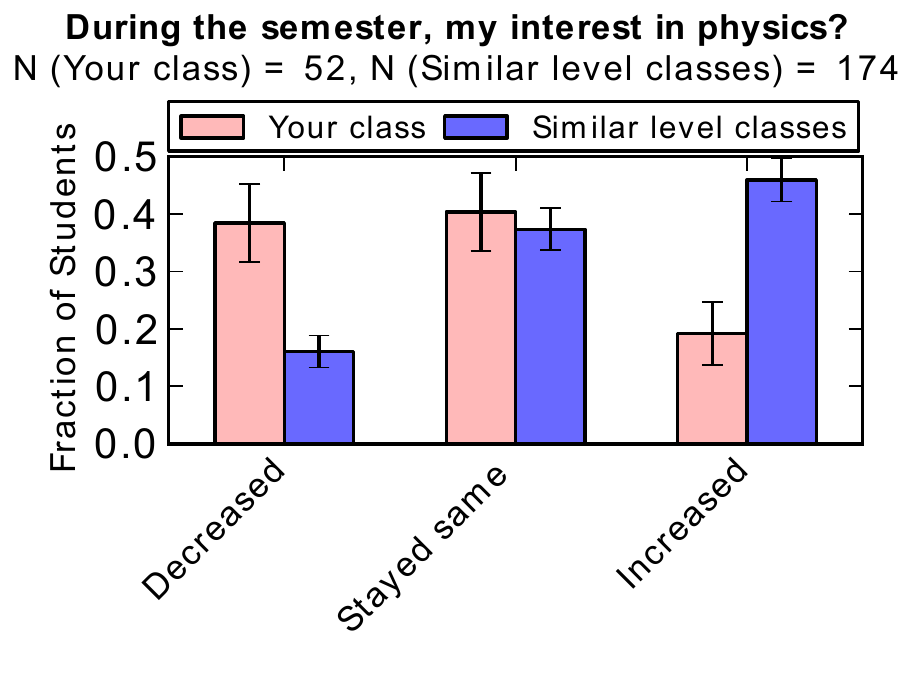}
\caption{Change in students' interest in physics.  \textbf{Your class} (Red) refers to your own class. \textbf{Similar level classes} (Blue) refers to a set of students in all classes at the calculus-based introductory-level.}
\label{fig:Change_in_interest}
\end{figure}

Finally, instructors are presented with basic demographic information about their class, which is obtained from a short appendix at the end of the post E-CLASS. Most importantly, instructors see the distribution of students' majors in their own class and in similar-level classes, so they can readily compare the composition of their class to others.  This is especially important for introductory courses that may target specific majors (e.g., non-sciences, life-sciences, or physical sciences and engineering). Also, instructors are provided with figures summarizing students' responses to ``Currently, what is your level of interest in physics?'' and to ``During the semester, my interest in physics (increased, decreased, or stayed the same).''  Figure\ \ref{fig:Change_in_interest} shows data about students' change in interest.

Currently, we know of six schools that are actively using the E-CLASS reports as an assessment tool for their curricula.  Four schools are using it for evaluation of significant curricular changes to their introductory lab sequence, 
while two others are using it for evaluations of upper-division laboratory courses.  
We are actively soliciting feedback from instructors on how to make our survey and reports more useful for course evaluation.  In response to feedback on the Fall 2012 E-CLASS reports, we now include: a summary table of class participation (Table\ \ref{tab:Participation_Summary}), an overall E-CLASS score for the entire class displayed as a bar graph (Fig.\ \ref{fig:Overall_ECLASS_score}), and the ``How important for earning a good grade...'' information is presented graphically rather than as a table (Fig.\ \ref{fig:Grade_plot}).  Additional input from instructors will allow us to further condense our reports and bring out the most salient features.  Our efforts to provide efficient and helpful information to faculty about their courses and to have this information promote changes in classroom instruction is a goal we share with other current projects such as Data Explorer and Assessment Resources for Faculty (DEAR-Faculty), which is an assessment-focused extension of the PER Users' Guide,\footnote{URL: http://perusersguide.org/} and the 2013 American Association of Universities Undergraduate STEM Education Initiative,\cite{AAU2013}
 which is focusing on overcoming challenges to adopting best teaching practices, including assessing student learning and assessment of classroom practices.
\section{Large-scale survey administration and participation}
\subsection{Participation}
During the Fall of 2012 and Spring of 2013, E-CLASS was administered to 45 different classes at 20 institutions in 3 countries. 
The institutions represent a wide cross-section of institution types (7 Private, 13 Public), sizes (5 Small (1-3K), 3 Medium (3-10K), 12 Large(10K+)), and degree-granting statuses (1 associates, 5 baccalaureate, 3 masters, 11 PhD).  The 45 individual classes included: 11 algebra-based introductory-level classes, 18 calculus-based introductory-level classes, and 16 laboratory classes beyond the intro-level, which were typically for physics and engineering physics majors.  The introductory classes tended to be larger, many in the range 50-200 students, while the upper-division classes were typically smaller, mostly in the range 8-25 students.  The median completion time on the Spring 2013 pre E-CLASS was 8 minutes ($N$=745), while for the post E-CLASS was 11 minutes ($N$=521).  The relatively short completion times are made possible by the reliance on pairs and triplets of questions around a single concept.  Further, the online administration allows the reading of the statement and the response to be immediately linked, which is an advantage over paper-based surveys that use ``bubble sheets'' for collecting responses.

\begin{figure}
\includegraphics[width=0.40\textwidth, clip, trim=0mm 0mm 0mm 0mm]{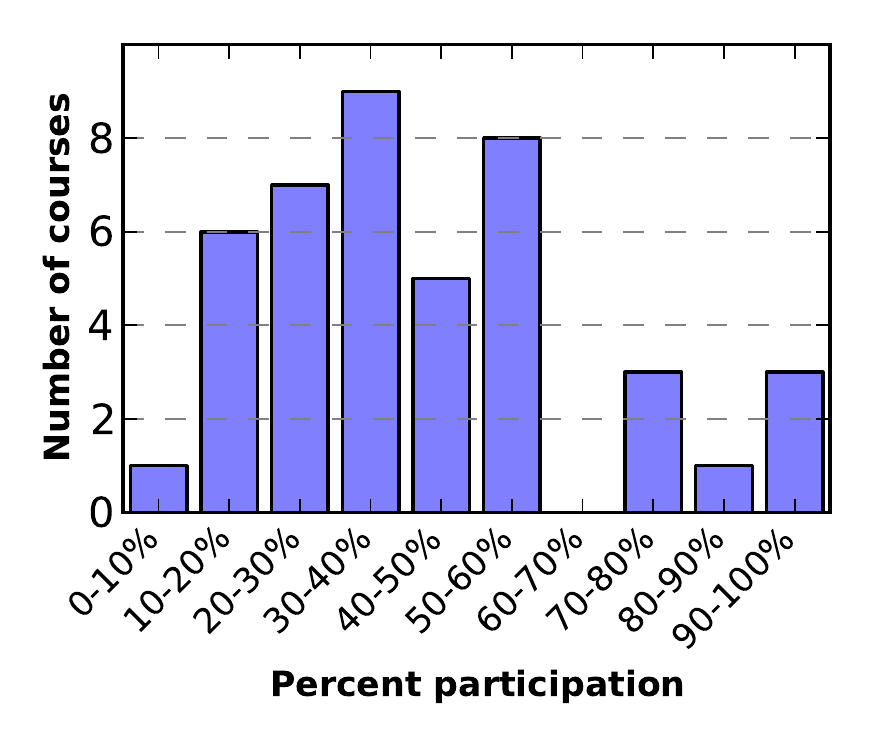}
\caption{Distribution of participation levels for all classes taking the E-CLASS survey.  The percentage is calculated by diving the number of matched pre/post responses by the total number of students reported to be enrolled in the class.}
\label{fig:ECLASS_participation}
\end{figure}

Although we received responses from a large number of institution and classes, the response rate in about half of those classes was disappointingly low.  Fig.\ \ref{fig:ECLASS_participation} shows the distribution of E-CLASS response rates for all 45 classes.  Only 20 of the 45 classes had a matching pre/post response rate higher than 40\%.  By comparison, when other surveys, such as CLASS-Phys, are routinely administered at CU for a small amount of credit and with multiple reminders from the instructor, the response rate is typically between 45\% and 60\%.    The lowest E-CLASS response rates occurred when faculty chose not to give any credit for completion of the survey, which is contrary to established recommendations for achieving high levels of participation.\cite{Adams2011}  


\subsection{Administration}
Delivering the survey online made it easy for instructors to adopt the E-CLASS in their classes.  However, the full administration of the survey was still highly labor intensive and required many steps for each individual class.  
Based on these experiences, future versions of the E-CLASS will likely be administered in a more unified online environment.  In this unified environment, instructors would be able to create an account for their class, enter basic information about their class and institution, get a unique survey link to send to their students,  have immediate access to lists of students completing the survey, and have immediate access to the aggregate E-CLASS report after the close date on the survey.  We hope that by providing an integrated environment for the survey and results, instructors will receive information in a timely manner, that the E-CLASS can more easily be integrated into courses, that students will respond at a higher rate, and that there will be fewer errors in selecting the appropriate course names and course sections.

\section{Conclusions}

The E-CLASS survey was motivated by the evident gap between common student practices in many laboratory courses and the epistemological beliefs, habits of mind, and scientific practices essential for successfully engaging in research.  The E-CLASS was developed as an epistemology and expectations survey to directly assess students views of doing physics experiments in both the classroom context and in the context of professional research.  Initial results show evidence of some significant gaps between students' epistemology of classroom experiments and research experiments (e.g., the role of asking questions when doing experiments).   Because evidence of validation has been gathered from a wide student population, the E-CLASS can be administered in any undergraduate physics lab, and to-date has received responses from 45 different laboratory classes at 20 institutions.  In order to demonstrate its use as a course assessment tool, partial results from the instructor report for a calculus-based physics lab at a large research university were presented.  On-going studies include a comparative evaluation of different laboratory curricula and the evaluation of laboratory activities in a massive open online course (MOOC).  Future work will discuss the curricular details of these lab-centered courses and the influence they may be having on students' epistemology.  As the administration and processing of results continues to be streamlined, we plan to provide access to any interested instructors nationally and internationally. 

\section{Acknowledgments}

The authors would like to thank the CU-Boulder Physics Department and Physics Education Research Group for contributing to the learning goals process and providing feedback on early versions of the survey.  We would also like to particularly thank the Advanced Lab Physics Association (ALPhA) community for their support in disseminating and providing feedback on a number of aspects of the E-CLASS development.  This work is supported by NSF CAREER PHY-0748742, NSF TUES DUE-1043028, JILA PFC PHY-0551010, the CU Science Education Initiative, and the Center for STEM Learning NSF DRL-0833364.  The views expressed in this paper do not necessarily reflect those of the National Science Foundation.

\appendix*\section{List of E-CLASS Statements}
\label{sec:List_of_statements}


\begin{longtable*}{|p{.02\textwidth}|p{.5\textwidth}|p{.35\textwidth}|p{.06\textwidth}|}

 \caption{List of all E-CLASS statements.  The personal and professional epistemology statements go with the pair of questions ``What do YOU think when doing experiments for class?'' and ``What would experimental physicists say about their research?''  The third column lists the expectation question that forms a triplet with the personal and professional epistemology question.  `NA' means no expectation question is associated with that particular epistemological construct. The final column gives the expert consensus (A = agree, D = disagree).  Question 23 is omitted because it is a check question to make sure students are reading the statements.}  \label{grid_mlmmh} \\

\hline
&                   \textbf{Personal and Professional Epistemology Statement} & \textbf{How important for earning a good grade in this class was... } & \textbf{Expert}\\\hline
\endfirsthead

\multicolumn{4}{c}%
{{\bfseries \tablename\ \thetable{} -- continued from previous page}} \\
\hline &                   \textbf{Personal and Professional Epistemology Statement} & \textbf{How important for earning a good grade in this class was... } & \textbf{Expert}\\\hline 
\endhead

 \multicolumn{4}{|r|}{{Continued on next page}} \\ \hline
\endfoot

\hline \hline
\endlastfoot

   1 &                                                                 When doing an experiment, I try to understand how the experimental setup works. &                                                     ...understanding how the experimental setup works?  & A\\ \hline
      2 &                                                                                      If I wanted to, I think I could be good at doing research. &                                                                                                      NA  & A\\\hline
      3 &                                                          When doing a physics experiment, I don't think much about sources of systematic error. &                                                         ...thinking about sources of systematic error?  & D\\\hline
      4 &                  If I am communicating results from an experiment, my main goal is to create a report with the correct sections and formatting. &                                     ...communicating results with the correct sections and formatting? & D\\\hline
  5 &                                                                        Calculating uncertainties usually helps me understand my results better. &                                          ...calculating uncertainties to better understand my results?  & A\\\hline
      6 &                                                Scientific journal articles are helpful for answering my own questions and designing experiments &                                                                ...reading scientific journal articles? & A\\\hline
      7 &                                                                                                       I don't enjoy doing physics experiments.  &                                                                                                      NA  & D\\\hline
      8 &                                                                           When doing an experiment, I try to understand the relevant equations. &                                                               ...understanding the relevant equations? & A\\\hline
      9 &                          When I approach a new piece of lab equipment, I feel confident I can learn how to use it well enough for my purposes.  &                                                ...learning to use a new piece of laboratory equipment?  & A\\\hline
     10 &                                                         Whenever I use a new measurement tool, I try to understand its performance limitations. &                                 ...understanding the performance limitations of the measurement tools?  & A\\\hline
     11 &                                                                                         Computers are helpful for plotting and analyzing data.  &                                                   ...using a computer for plotting and analyzing data? & A \\\hline
     12 &                                      I don't need to understand how the measurement tools and sensors work in order to carry out an experiment. &                                           ...understanding how the measurement tools and sensors work?  & D\\\hline
     13 &                                                                                If I try hard enough I can succeed at doing physics experiments. &                                                                                                      NA & A \\\hline
     14 &                                                                   When doing an experiment I usually think up my own questions to investigate.  &                                                        ...thinking up my own questions to investigate?  & A\\\hline
     15 &                                                               Designing and building things is an important part of doing physics experiments.  &                                                                      ...designing and building things?  & A\\\hline
     16 &                                                       The primary purpose of doing a physics experiment is to confirm previously known results. &                                                                 ...confirming previously known results? & D\\\hline
     17 &                                              When I encounter difficulties in the lab, my first step is to ask an expert, like the instructor.  &                                              ...overcoming difficulties without the instructor's help? & D\\\hline
     18 &                                                     Communicating scientific results to peers is a valuable part of doing physics experiments.  &                                                          ...communicating scientific results to peers?  & A\\\hline
     19 &                                                                           Working in a group is an important part of doing physics experiments. &                                                                                 ...working in a group? & A\\\hline
     20 &                                                                                             I enjoy building things and working with my hands.  &                                                                                                      NA & A\\\hline
     21 &  I am usually able to complete an experiment without understanding the equations and physics ideas that describe the system I am investigating. &          ...understanding the equations and physics ideas that describe the system I am investigating?  &D \\\hline
     22 &              If I am communicating results from an experiment, my main goal is to make conclusions based on my data using scientific reasoning. &                                        ...making conclusions based on data using scientific reasoning?  & A\\\hline
     24 &                                                  When I am doing an experiment, I try to make predictions to see if my results are reasonable.  &                                             ...making predictions to see if my results are reasonable? & A \\\hline
     25 &                                                               Nearly all students are capable of doing a physics experiment if they work at it. &                                                                                                      NA & A\\\hline
     26 &                             A common approach for fixing a problem with an experiment is to randomly change things until the problem goes away. &                                      ...randomly changing things to fix a problem with the experiment? & D\\\hline
     27 &                                                                    It is helpful to understand the assumptions that go into making predictions. &  ...understanding the approximations and simplifications that are included in theoretical predictions? & A \\\hline
     28 &                                                  When doing an experiment, I just follow the instructions without thinking about their purpose. &                                    ...thinking about the purpose of the instructions in the lab guide? &  D\\\hline
     29 &                                                                        I do not expect doing an experiment to help my understanding of physics. &                                                                                                      NA & D\\\hline
     30 &                                If I don't have clear directions for analyzing data, I am not sure how to choose an appropriate analysis method. &                     ...choosing an appropriate method for analyzing data (without explicit direction)? & D\\\hline
     31 &                                                                           Physics experiments contribute to the growth of scientific knowledge. &                                                                                                      NA & A\\\hline

\end{longtable*}

\bibliography{C:/SugarSync/Ben/Bibliographies/Publications-PRST_PER_ECLASS_Intro}

\end{document}